\documentclass[12pt]{amsart}
\usepackage[top=0.85in,left=1.2in,right=1.2in,footskip=0.75in]{geometry}

\usepackage{graphicx}
\usepackage{multirow}
\usepackage{url}
\usepackage{amsmath,amssymb}

\newcommand{\vY}{\boldsymbol{Y}}

\newcommand{\NN}{\mathbb{N}}
\newcommand{\ZZ}{\mathbb{Z}}

\newcommand{\RR}{\mathbb{R}}

\newcommand{\ud}{\textup{d}}

\begin{document}

\title[Wave-shape functions and pulse signals]{Modeling the pulse signal by wave-shape function and analyzing by synchrosqueezing transform}

\author[]{Hau-Tieng~Wu\textsuperscript{1},
Han-Kuei~Wu\textsuperscript{2,3},
Chun-Li~Wang\textsuperscript{4},
Yueh-Lung Yang\textsuperscript{5},
Wen-Hsiang~Wu\textsuperscript{6},
Tung-Hu~Tsai\textsuperscript{2}
and Hen-Hong~Chang\textsuperscript{7,8}}

\address{(1) Mathematics, University of Toronto, Toronto, Ontario, Canada.}
\address{(2) Institute of Traditional Medicine, School of Medicine, National Yang-Ming University, Taipei, Taiwan.}
\address{(3) Department of Chinese Medicine, Taiwan Landseed Hospital, Tao-Yuan, Taiwan.}
\address{(4) Cardiovascular Division, Department of Internal Medicine, Chang Gung Memorial Hospital, Taoyuan, Taiwan. College of Medicine, Chang Gung University, Taiyuan, Taiwan.}
\address{(5) Center for Traditional Chinese Medicine, Keelung Chang Gung Memorial Hospital, Taiwan.}
\address{(6) Department of Healthcare Management, Yuanpei University of Medical Technology, Hsinchu, Taiwan.}
\address{(7) School of Post-Baccalaureate Traditional Chinese Medicine, and Research Center for Chinese Medicine and Acupuncture, China Medical University, Taiwan.}
\address{(8) Departments of Chinese Medicine, China Medical University Hospital, Taiwan.}

\maketitle

\begin{abstract}
We apply the recently developed adaptive non-harmonic model based on the wave-shape function, as well as the time-frequency analysis tool called synchrosqueezing transform (SST) to model and analyze oscillatory physiological signals. To demonstrate how the model and algorithm work, we apply them to study the pulse wave signal. By extracting features called the spectral pulse signature, {and} based on functional regression, we characterize the hemodynamics from the radial pulse wave signals recorded by the sphygmomanometer. 
Analysis results suggest the potential of the proposed signal processing approach to extract health-related hemodynamics features. 
\end{abstract}

\section{Introduction}
Hemodynamics is an essential ingredient of cardiovascular physiology which not only reflects the forces the heart needs to pump blood through the cardiovascular system, but also reflects the integrity of {an individual's} physiological system. While there are {many} aspects {to} hemodynamics \cite{Milnor:1989,Nicholas_ORourke:2005,Nicholas_ORourke_Vlachopoulos:2011,Caro_Pedley_Schroter_Seed:2011,Salvi:2012}, it is common to evaluate hemodynamics by assessing the pulse wave signals recorded {at} different locations. For example, the brachial cuff systolic and diastolic blood pressures derived from the pulse wave {have important clinical applications,  long established beginning} 1870 \cite{Lewington_Clarke_Qizilbash_Peto_Collins:2002}. 
In addition to the blood pressure, the pulse wave signal itself contains {an abundance of} clinical information. For example, {in} subjects with dysfunctional left ventricle, different carotid pulse patterns {can be observed, such as} hyperkinetic pulse, pulsus alternans, pulse bisferiens, pulsus parvus et tardus, etc \cite{Morris:1990,Braunwald_Perloff:2014}.  \footnote{The {\em hyperkinetic pulse} is characterized by an increase in the velocity of the upstroke and amplitude. The {\em pulsus alternans} is a beat-to-beat variation in the amplitude of the pressure pulse. The {\em pulsus bisferiens} arterial pulse is perceived as two narrowly separated positive waves during systole. The {\em pulsus pervious et tardus} is a small and delayed arterial pulse.}
However, it is difficult to {obtain these} carotid pulse wave signals non-invasively \cite{Chen_Ting_Nussbacher_Nevo_Kass_Pak_Wang_Chang_Win:1996}. On the other hand, the non-invasive pulse diagnosis (pulsology) based on, for example, the brachial or radial pulse signal, provides several {aspects of} physiological information, {such as} the central pressure wave \cite{Scuteri_Chen_Yin_Chih_Spurgeon_Lakatta:2001,ORourke_Pauca_Jiang:2001}. 
Other indices associated with the pulse wave signals recorded from different locations, {including} the augmentation index \cite[Section 6.1]{Salvi:2012}\footnote{The augmentation index indicates the incidence of reflected waves on the total pulse pressure. See \cite[Section 6.1]{Salvi:2012} for a definition.} 
\cite{Chen_Ting_Nussbacher_Nevo_Kass_Pak_Wang_Chang_Win:1996,Vlachopoulos_Aznaouridis_ORourke_Safar_Baou_Stefanadis:2010}, are {also} available for clinical {applications. Since} there is a {great amount of} information {within} the pulse signal, a {deeper and more extensive understanding} of the pulse wave is undoubtedly important to {better} assess not only the cardiovascular but also the physiological integrity. 

The common {approaches to analyzing} the pulse wave signals {can} be classified into two categories -- time domain analysis {and} frequency domain analysis. In both categories, {it is necessary to have either} a representative pulse wave cycle, which reflects the interaction between the one time heart pump driving force and the impedance to blood flow, or a sequence of pulse wave cycles over a period of time, during which the heart rate is relatively constant. 
In the time domain analysis, different landmarks or features are identified from the pulse signal. For example, {with the} radial pulse signal {is characterized by} 
the height of the percussive wave, the height of the dicrotic notch, the length of the cardiac cycle, the length of acute ejection, etc.
Different features with different physiological meanings {can} be obtained from the pulse signals recorded from other areas, and these features have been widely used in clinics for health evaluation \cite{Braunwald_Perloff:2014}. 
{Furthermore, new mathematical methods have been applied to investigate additional physiological information. For example, the central pulse wave could be reconstructed from the radial pulse wave by a generalized transfer function, and then be separated into inflection and reflection waves in order to evaluate markers of early vascular aging \cite{Melenovsky_Borlaug_Fetics_Kessler_Shively_Kass:2007}.}
In the frequency domain analysis, the energy of different harmonic modes estimated by power spectrum analysis are {also} potential features for clinical {applications}. According to the resonance theory \cite{Wang_Chang_Wu_Hsu_Wang:1991}, different harmonic modes are associated with the integrity of different organs \cite{Hsu_Chao_Hsiu_Wang_Li_Wang:2006}.
Several existing works show the potential of this spectral approach, {such as} the relationship between the internal organ disorders and the spectral content \cite{Wei_Lee_Chow:1984,Huang_Wei_Liao_Chang_Kao_Li:2011}. {Although} this approach is commonly applied, {it has} two significant limitations. 

\begin{figure}[h!]
    \begin{center}
    \includegraphics[width=.75\textwidth]{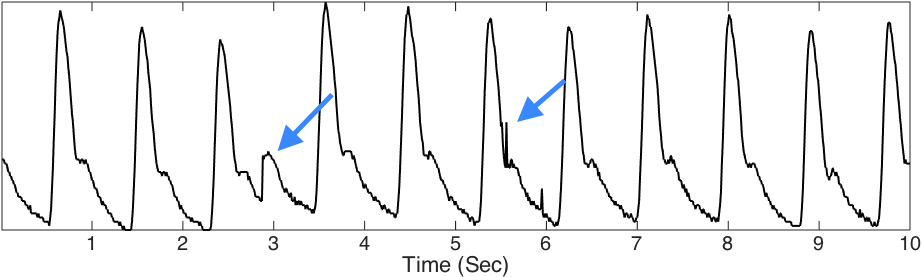}     
    \end{center}
    \caption{An example of the radial pulse signal recorded from a subject with congestive heart failure. Note that at the $3^{rd}$ second there is an obvious artifact, indicated by the blue arrow, {and} at the $5.5^{th}$ second there are two shock noise events. The relatively small pulse wave at the $3^{rd}$ second is related to a {premature }atrial contraction. {This is significant since patients} with heart failure may have more premature beats than a normal control.}\label{fig:PulseExample}
  \end{figure}

The first limitation is the over-simplification of the model underlying the analysis techniques. For example, heart rate variability (HRV) and pulsus alternans are often {not used} in the analysis. {Though} the time domain landmark features reflect the hemodynamics from different aspects, they might be confounded by HRV and pulsus alternans. Since HRV and pulsus alternans are inevitable physiological dynamics, {without considering them in the model and analysis, both time domain and frequency domain analysis results may be inaccurate and the results might lead to incorrect interpretation.}
The second limitation is that it is not always straightforward to determine a representative pulse cycle or a period of time {when} the pulse wave signal is suitable for {these} analysis techniques, {so that} human intervention and subjective decision making are required. Indeed, to determine a pulse cycle, {it is necessary} to define what a pulse cycle means or, at least, {to determine} landmarks within the pattern to separate one pulse {cycle from} another. For example, {with an} electrocardiogram (ECG) signal, the repeating basic pattern is related to the electrophysiological activity of a normal heart beat, and a common chosen landmark {for this} is the R peak. As the R peak is usually dominant and significant, determining the oscillatory pattern for an ECG signal is not difficult. However, for pulse signal {such as those} acquired {in this study}, it is not always easy to define an oscillatory pattern or a landmark, in particular when the signal is recorded from an abnormal subject. Even {more seriously}, due to the {possibility of a} noisy record, the pulse cycle morphology {can} dramatically change from one pulse to another or {from} one subject to another (See Figure \ref{fig:PulseExample}). Although this difficulty can be mitigated to some extent by noise reduction algorithms, due to the non-linear nature of the signal, {their} performance is not guaranteed.

To address these limitations, in this paper we apply a recently proposed and analyzed descriptive model, called the {\em adaptive non-harmonic model}, to describe the pulse wave signals. {This} model is characterized by {what is referred to as the} {\em wave-shape function} \cite{Wu:2013}, which is defined {in order} to capture {what} a pulse wave cycle should look like, {including} the features in the time domain and frequency domain mentioned above. 
The main characteristic of the proposed model is that the HRV and pulsus alternans are decoupled from the wave-shape function. A companion algorithm, referred to as the {\em synchrosqueezing transform} (SST), is applied to provide an accurate estimation of the wave-shape function. SST is a recently proposed time-frequency analysis technique {that has a} mathematical foundation {and robustness to different kinds of noise and different practical issues have} been well established in the literature \cite{Daubechies_Lu_Wu:2011,Chen_Cheng_Wu:2014}. {By the adaptive non-harmonic model and SST, the above-mentioned limitations can be alleviated.} By a suitable manipulation of the estimated pulse wave cycle, a vector {can be obtained} as an associated feature, which further leads to our final index, {referred to as the} {\it global pulse signature (GPS)} \cite{Wu:2013,Chui_Lin_Wu:2016}. 

This paper is organized {as follows}. In Section \ref{Section:methodology}, we systematically model the pulse signal {using} the notion of wave-shape function and the adaptive non-harmonic model. In Section \ref{Section:MaterialandMethod}, we summarize the SST algorithm and discuss how the wave-shape function {can be estimated} based on the SST and functional regression technique and {then} generate the GPS index. In Section \ref{Section:materialresult}, we apply the proposed algorithm to study the radial pulse signal. The data collection procedure, the physiological background of pulsology and the analysis result are reported. In Section \ref{Section:discussion}, we discuss our findings {with their} clinical interpretations and indicate {directions for} future {research}.

\section{Adaptive Non-Harmonic Model}\label{Section:methodology}

\subsection{Adaptive non-harmonic model for the pulse signal}\label{Section:AdaptiveHarmonicModel}

In this section, we propose {a} mathematical model to describe the pulse wave signal. It is well known that hemodynamics is a {highly complex} subject \cite{Nicholas_ORourke:2005,Nicholas_ORourke_Vlachopoulos:2011,Caro_Pedley_Schroter_Seed:2011,McEniery_Cockcroft_Roman_Franklin_Wilkinson:2014,Westerhof_vanderWijngaard_Murgo_Westerhof:2008}. 
Different signals recorded from different areas of the body by different instruments {can shed light on} different aspects of hemodynamics. 
The question of modeling hemodynamics has been discussed by several authors, {who have proposed} different models based on different physiological factors, {such as} pressure gradient, resonance, vibration, etc, have been proposed; see \cite{Milnor:1989,Wang_Hsu_Jan_Wang:2010,Nicholas_ORourke_Vlachopoulos:2011,Caro_Pedley_Schroter_Seed:2011} and the literatures cited {therein} for detailed discussions {of} these models. While these models shade a light {on} the understanding of the pulse wave signal, {having} a signal model to {effectively} quantify the pulse behavior {remains a distant goal}. 
In particular, inevitable physiological {aspects, such as} HRV and pulsus alternans, are commonly missed or {disregarded} in the existing models. Ignoring them in the model and analysis, {however}, might not only limit the analysis but also lead to {incorrect} findings. In this paper, we apply a different model and approach to {resolve} these issues. 

The pulse wave signal is oscillatory in nature, and there are several well-known features {that we consider here}: how fast the signal oscillates, how much the oscillatory rate changes, how {differently} the signal behaves inside an oscillation, etc. Understanding these features plays a significant role in {the} data analysis, which {can help} to obtain more information about the {physiological} system. Based on these observations, we consider the following adaptive non-harmonic model\cite{Wu:2013} to model the pulse signal. This model {differs} from most existing models in that it is purely {\em phenomenological}; that is, the parameters in the model are solely driven by observations of the physiological signal, {rather than by} explicit and quantitative underlying mechanisms. 

\subsubsection{Wave shape function}
Based on the periodic nature of an oscillatory function, we introduce the {\it wave shape function}, which is {used} to model how a signal oscillates over the period. This idea has been {previously} studied in \cite{Wu:2013,Yang:2014,Hou_Shi:2016}.
{To resolve the difficulty of defining a period}, we shall call a function $f$ {\it $\tau-$periodic}, where $\tau>0$, if (1) the periodic condition is satisfied; that is, for all $t \in \RR$ and all $k \in \ZZ$, $f(t+k\tau)=f(t)$; and (2) for all $0<\tau'<\tau$ the periodic condition is not satisfied.
Take two positive values $\delta$ and $\theta$ and a positive integer $D$. A $1$-periodic function $s$ is called a {\it wave shape function} with parameters $\delta$, $D$ and $\theta$ if the following conditions are satisfied. First, {the function} is differentiable and its derivative is a Holder continuous function with the Holder coefficient $\alpha>1/2$. 
Denote $\widehat{s}$ to be the Fourier transform of the function $s$.
The second condition {that} a wave shape function should satisfy is that $s$ has the unit energy (unit $L^2$-norm) and all the Fourier modes $\widehat{s}(k)$, $k\neq 1$ are dominated by the product of $\delta$ and the first mode coefficient; that is,
\begin{equation}\label{definition:shape:1}
\forall k \in \NN, \, \mbox{ with }k\neq 1, \, \left| \widehat{s}(k) \right| \leq \delta \, \left| \widehat{s}(1) \right|.
\end{equation}
Furthermore, {it is desirable for} $\widehat{s}$ {to be} mostly concentrated in the low frequency region, which is quantified by $\theta$ by:
\begin{equation}\label{definition:shape:2}
\sum_{n>D}|n\widehat{s}(n)|\leq \theta.
\end{equation}
{Conditions} (\ref{definition:shape:1}) and (\ref{definition:shape:2}) {reflect} the practical finding of the spectral analysis of the pulse signals \cite{Taylor:1966,Wei_Chow:1985,LinWang_Jan_Shyu_Chiang_Wang:2004}, {and} it {can be} observed that the amplitudes of tenth and higher order harmonics are negligible. 
Note that the commonly selected landmarks and lengths considered in the time domain analysis could be understood as morphological features describing the wave-shape function {that is} modeling how a pulse repeats itself. While there is a one-to-one relationship between these landmarks and Fourier coefficients $\widehat{s}$, this relationship is nonlinear in nature. Thus, while we could view the Fourier coefficients of the wave shape function as the ``features'' of the pulse wave,  their physiological interpretations are not directly related to the {hemodynamic} interpretations of these landmarks. See Figure \ref{fig:ShapeExample} for an example of the wave-shape function describing the pulse wave signal.

\subsubsection{Adaptive non-harmonic model}
{Using} the notion of wave shape function, we now describe our phenomenological model to capture the recorded pulse signal.
Fix parameters $\delta,\,\theta,\,D$ for a wave shape function $s$. We consider the following \textit{Intrinsic Mode Type (IMT) functions} to model the pulse wave signal. An IMF function, $f$, is a bounded and continuous function with a continuous derivative {that} satisfies the following format:
\begin{equation}\label{definition:IMT}
f(t)=A(t)s(2\pi \phi(t)),
\end{equation}
where $A$ is a positive differentiable function and $\phi$ is a monotonically differentiable function. Intuitively, $A(t)$ describes how large the oscillation is at time $t$, and the positive function $\phi'(t)$, the first derivative of $\phi$, describes how fast the oscillation is at time $t$. To see how it is interpreted in this way, consider a constant positive function $A$ and a linear function with a positive slope as $\phi$. Suppose $\phi(t)=\xi_0 t$, where $\xi_0>0$. In this case, we know that $f$ is a harmonic function with the frequency $\xi_0$ and the amplitude $A$.  We could thus view (\ref{definition:IMT}) as a generalization of the harmonic function. 
{Though} the heart rate is not constant, normally it does not change suddenly. {Therefore} we need following conditions to better quantify the pulse signal. Fix a small positive constant $\epsilon$. Then we assume that $A(t)$ does not change too fast in the sense that its derivative is bounded by $\epsilon \phi'(t)$ and that $\phi'(t)$ does not change too fast in the sense that $\phi''(t)$ exists and is bounded by $\epsilon\phi'(t)$; that is, we have 
\begin{equation}\label{definition:IMT2}
|A'(t)|\leq \epsilon\phi'(t)\mbox{ and }|\phi''(t)|\leq \epsilon\phi'(t)\mbox{ for all time }t.
\end{equation} 
We would thus model the pulse wave signal by (\ref{definition:IMT}) with the condition (\ref{definition:IMT2}), and call this model {the} {\it adaptive non-harmonic model}.

We would call the monotonically increasing function $\phi(t)$ the {\it phase function}, $\phi'(t)$ the {\it instantaneous frequency} (IF), and $A(t)$ the {\it amplitude modulation} (AM). 
An important issue regarding the identifiability issue of the phase function, the IF and the AM {cannot} be ignored if the {discussion is to be} rigorous. Indeed, {there might be} infinitely many different ways to represent a cosine function $g_0(t)=\cos(2\pi t)$ in the format $a(t)\cos(2\pi\phi(t))$ so that $a>0$ and $\phi'>0$, even though it is well known that $g_0(t)$ is a harmonic function with amplitude $1$ and frequency $1$. Indeed, we could find infinitely many smooth functions $\alpha$ and $\beta$ so that $g_0(t)=\cos(t)=(1+\alpha(t))\cos(2\pi(t+\beta(t)))$, and in general there is no reason to favor $\alpha(t)=\beta(t)=0$. Clearly, before resolving this issue, {amplitude $1$ and frequency $1$ cannot be taken as} reliable features to quantify the signal $g_0$. This identifiability issue has been well studied in \cite{Chen_Cheng_Wu:2014,Kowalski_Meynard_Wu:2015}, {finding} that if $g_0(t)=A(t)s(\phi(t))$ and $g_0(t)=(A(t)+\alpha(t))s(\phi(t)+\beta(t))$ also holds, and both satisfy the condition (\ref{definition:IMT2}), then $|\alpha(t)|$ and $|\beta'(t)|$ are both bounded by $\epsilon$ for all time $t\in\RR$. Note that the IF and AM are always positive, but usually not constant. Clearly, when $\phi$ is a linear function with a positive slope and $a$ is a positive constant, then the model is reduced to the harmonic model and the IF is equivalent to the notion frequency in the ordinary Fourier transform sense. The conditions $|a'(t)|\leq \epsilon \phi'(t)$ and $|\phi''(t)|\leq \epsilon\phi'(t)$ force the signal to {locally oscillate ``regularly''}, that is, around time $t_0$, the signal $A(t)s(\phi(t))$ oscillates like $A(t_0)s(\phi(t_0)-t_0\phi'(t_0)+\phi'(t_0)t)$, and hence the nominations of IF and AM. 
We mention that {this} model is a special case of a wider class of model composed of multiple oscillatory components in the sense that we only have one oscillatory component in the model. 
For details about a more general model, see \cite{Daubechies_Lu_Wu:2011,Wu:2013,Chen_Cheng_Wu:2014,Kowalski_Meynard_Wu:2015}.

\subsubsection{Physiological interpretation}

The main reason we consider a time-varying frequency and amplitude model like (\ref{definition:IMT}) is to capture the physiological facts {of} HRV and pulsus alternans. 

{Since} the heart does not beat at a constant rate \cite{Malik_Camm:1995}, the pulse signal should not oscillate with a constant frequency. The non-constant heart rate is modeled by IF and hence the HRV could be evaluated from analyzing IF. In the past decade, due to the health information {embedding} in HRV and the trend {towards} personalized health care {requirements}, estimating the HRV from the pulse signal extracted from different resources has attracted {significant research} \cite{Gil_Orini_Bailon_Vergara_Mainardi_Laguna:2010,Davila:2012Thesis,Wu_Lewis_Davila_Daubechies_Porges:2015}. However, the existence of HRV is commonly ignored in the pulse analysis literature. For example, in the traditional spectral analysis approach to the radial pulse, {researchers need} to analyze the pulse signal over the interval where the heart rate {closely resembles} a constant \cite{LinWang_Jan_Shyu_Chiang_Wang:2004,Hsu_Chao_Hsiu_Wang_Li_Wang:2006}. 

\begin{figure}[h!]
    \begin{center}
    \includegraphics[width=.75\textwidth]{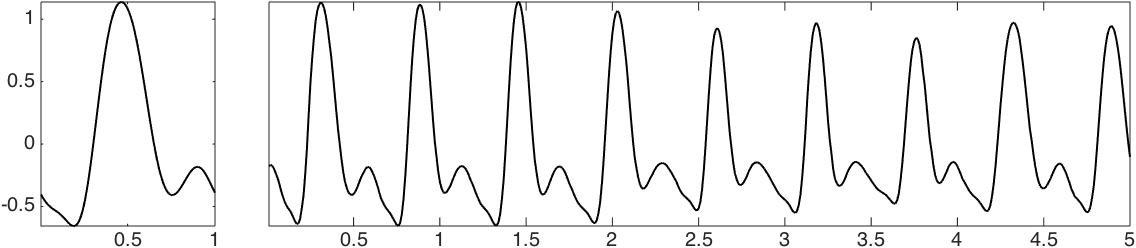}     
    \end{center}
    \caption{Left: an illustrative wave-shape function $s$ with $D=5$, $\delta=0.59$ and $\theta=0$. Right: an IMT function with the wave-shape function $s$ and time varying amplitude and frequency. Here the frequency is about $1.2$ Hz.}\label{fig:ShapeExample} \end{figure}

The possible pulsus alternans is captured by the AM. This model is particularly important in {patients susceptible to} heart failure {since} pulsus alternans is a manifestation of severe impairment of the left ventricular systolic function \cite{Morris:1990}.
An example of a wave-shape function and {an} IMT function representing the pulse signal {are} shown in Figure \ref{fig:ShapeExample}.

\subsubsection{Recorded pulse signal}
In practice, noise is inevitable {when recording} signals. Thus, we model the recorded pulse signal by
\begin{align}\label{decompShape}
Y(t)=A(t)s(\phi(t))+\sigma(t)\Phi(t),
\end{align}
where $A(t)s(\phi(t))$ is the adaptive non-hamornic model for the pulse signal, $\sigma$ is a slowly varying smooth function quantifying the possible non-stationarity in the data collection process, and $\Phi$ is a stationary random process with unit standard deviation describing the noise. See, for example, \cite{Chen_Cheng_Wu:2014} for a discussion of the noise. Here, we do not assume that the noise is Gaussian or white, while the commonly encountered Gaussian white noise is a specific example. Since in general the noise might be complicated with time-dependence and {be} far from Gaussian, we take this noise model into account. 

While the pulse signal could be recorded by different ways, like tonometer, photoplethymography, video \cite{Davila:2012Thesis}, etc, the wave-shape function could be different for different types of pulse signals.

\section{Methodology}\label{Section:MaterialandMethod}

\subsection{Synchrosqueezing transform}

It is a common practice to apply {the} Fourier transform to study oscillatory signals, in particular the pulse wave signal {in which we are interested} \cite{Wei_Lee_Chow:1984,Hsu_Chao_Hsiu_Wang_Li_Wang:2006,Huang_Wei_Liao_Chang_Kao_Li:2011}. As useful as the spectral analysis is, however, it {is} well known that when the signal is not composed of harmonic functions, the power spectrum determined by the Fourier transform might be misleading. {For} spectral pulse wave analysis, since HRV and pulsus alternans are inevitable, in practice analysts {must} carefully choose the signal. {But} there is {still} no guarantee that the HRV and pulsus alternans influence could be eliminated, and {this could} become a confounder in the analysis. 

While the oscillatory signals with time-varying AM and IF are ubiquitous, {much effort has been extended over} the past few decades to address this problem, {and several approaches {have been} proposed, ranging from empirical mode decomposition and its variations \cite{Huang_Shen_Long_Wu_Shih_Zheng_Yen_Tung_Liu:1998,Tavallali_Hou_Shi:2014,Cicone_Liu_Zhou:2014}, {to} optimized window and the approximation theory approach \cite{Ricaud_Stempfel_Torresani:2014,Chui_Mhaskar:2016}, to name but a few. {Moreover, research interest} has {recently} been extended to multivariate time series analysis \cite{Gao_Jin:2012,Gao_Yang_Fang_Jin_Xia_Hu:2015,Gao_Fang_Ding_Jin:2015,Ahrabian_Looney_Stankovic_Mandic:2015,Gao_Yang_Zhai_Ding_Jin:2016} to further take the spatial information into account. Among {the} different approaches, one} active field regarding this issue is time-frequency (TF) analysis. Based on different {principles}, several TF analysis techniques have been proposed. 
Typical examples include linear {methods such as} the short time Fourier transform (STFT) and {the} continuous wavelet transform (CWT){, or} the quadratic {methods such as} the Wigner-Ville distribution {or the Cohen class. We refer the reader to {a} standard textbook \cite{Flandrin:1999} for more information}. While these methods have attracted a {great deal} of attention in different fields other than signal processing, they are limited by {either} the Heisenberg {uncertainty} principal or the mixing issue. 
{We demonstrate this idea by discussing STFT}. The main idea {forming the basis for} STFT is dividing the signal into overlapping small pieces, and {then studying} the spectral behaviors over these small pieces. Mathematically, for a function $f$, its STFT associated with a window function $h(t)$ can be defined as
\[
V_f^{(h)}(t,\eta):=\int f(s)h(t-s)e^{-i2\pi \eta (t-s)}\ud s
\]
where $t\in\RR$ is the time, $\eta\in \RR^+$ is the frequency, $h$ is the window function determined by the user -- {for which} a common choice is the Gaussian function with kernel bandwidth $\sigma>0$, i.e. $h(t)=(2\pi\sigma)^{-1/2}e^{-t^2/\sigma^2}$. However, the Heisenberg uncertainty principal limits how well the spectrum {could be estimated} over each piece, {thereby limiting} the STFT. Similar discussions hold for CWT and other {linear} TF analysis techniques, and we refer the {reader} to \cite{Flandrin:1999} for details. {In the current work, in order to capture the hemodynamics, which have oscillations on the order of 1 second, we run STFT with $\sigma=0.5$ so that the window is not too long to lose the dynamic information, and is not too short to cause a numerical artifact.}

To handle these fundamental issues, {a nonlinear TF analysis technique,  synchrosqueezing transform (SST) \cite{Daubechies_Lu_Wu:2011,Chen_Cheng_Wu:2014}, which is a special reassignment technique (RM) \cite{Auger_Flandrin:1995}, is proposed to obtain a sharper time-frequency representation.}
Since {these techniques} were introduced, they have been widely applied in different fields; see \cite{Daubechies_Wang_Wu:2016} for a summary of its applications in different fields. 

Here we {briefly} summarize SST, and refer the reader to \cite{Auger_Flandrin:1995} for RM. 
{We mention that the SST could be applied to different linear TF analysis, like STFT, CWT, wave packet transform \cite{Yang:2014} or S-transform \cite{Huang_Zhang_Zhao_Sun:2015} and theoretically the results do not depend {significantly} on the chosen linear TF analysis. But to simplify the discussion, we choose to do the analysis with STFT.}
The Matlab code for the SST algorithm based on both CWT and STFT {can} be downloaded from \url{https://sites.google.com/site/hautiengwu/home/download}. 
In {brief}, the SST sharpens the TF representation determined by STFT by reallocating the STFT coefficients along the frequency axis to the ``correct'' frequency slot which represents the IF of an oscillatory component. 
Mathematically, The SST of $f$ is defined as
\[
S_f^{(h)}(t,\xi):=\lim_{\alpha \rightarrow 0} \int V_f^{(h)}(t,\eta)\, g_\alpha(|\xi-\omega^{(h)}_f(t,\eta)|)\, \ud \eta,
\]
where $g_\alpha$  is an approximate $\delta$-function in the sense that $g$ is {a} fast decaying smooth function with $\int g(x)\ud x =1$, so that $g_\alpha(t):=\frac{1}{\alpha}g(\frac{t}{\alpha})$ tends to the Dirac delta measure $\delta$ supported at $0$ weakly as $\alpha\to 0$, and with $\omega^{(h)}_f$ defined by
\[
\omega^{(h)}_f(t,\eta):= \left\{
\begin{array}{ll}
\frac{-i\partial_t V_f^{h}(t,\eta)}{2\pi V_f^{(h)}(t,\eta)} & \mbox{ if }\, V_f^{(h)}(t,\eta)\neq 0\\
-\infty & \mbox{ otherwise}\nonumber.
\end{array}\right.
\]
In plain language, by reading $\omega^{(h)}_f(t,\eta)$, we collect all STFT coefficients indicating the existence of an oscillatory component with frequency $\xi$ to the slot $\xi$. 
Note that {compared with RM,} in SST, the coefficients are reallocated along the frequency axis, so the causality is preserved; second, in SST we reallocate the STFT coefficient instead of the spectrogram coefficient. These two facts allow us to reconstruct the oscillatory components {of} interest, in particular when the signal is noisy or is composed of several oscillatory components. 
It has been {clearly found} that for the phenomenological model of the pulse wave signal {of} interest, the IF $\phi'(t)$ and the AM $A(t)$ {can be accurately estimated} from the recorded pulse signal \cite{Daubechies_Lu_Wu:2011,Chen_Cheng_Wu:2014}. Precisely,
{we} could prove that at time $t$, the coefficients in $S_f^{(h)}(t,\xi)$ are dominant when $\xi\approx\phi'(t)$. This property allows us an accurate estimate of the IF $\phi'$ by, for example, the curve extraction technique. Denote the estimated $\phi'$ by $\widetilde{\phi}'$.
We can then estimate the amplitude modulation $A(t)$ and the phase function $\phi(t)$ by building
\begin{align}
& \widetilde{R}(t):=h(0)^{-1}\int_{\{\xi:~|\widetilde{\phi}'(t)-\xi|\leq \epsilon^{1/3}\}}S_f^{(h)}(t,\xi)\ud \xi. \label{alogithm:sst:reconstruction}
\end{align}
The estimator of $A(t)$ is thus defined as $\widetilde{A}(t):=|\widetilde{R}(t)|$, and hence an estimator for $\phi(t)$, denoted as $\widetilde{\phi}(t)$, can be obtained by unwrapping the phase of the complex-valued signal $\widetilde{R}(t)/\widetilde{A}(t)$.
We refer {readers interested} in SST to \cite{Daubechies_Lu_Wu:2011,Chen_Cheng_Wu:2014} for the {detailed} numerical {algorithms and} the theory beyond {them}. 

\begin{figure}[h!]
    \begin{center}
    \includegraphics[width=.6\textwidth]{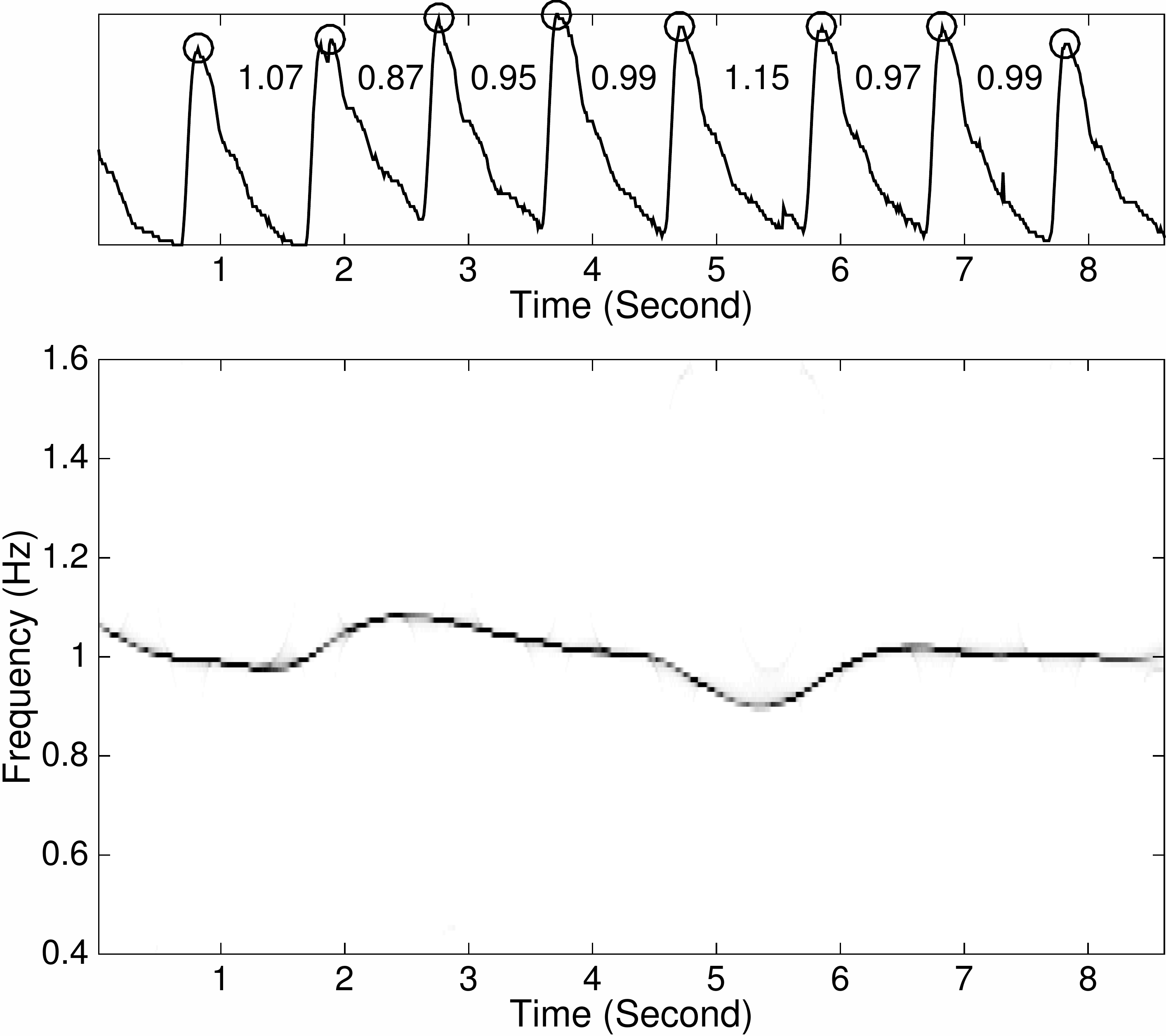}     
    \end{center}
    \caption{(a) The radial pulse signal $R(t)$ recorded from a normal subject. It is also clear that the IF and amplitude modulation are not constant due to the inevitable HRV and pulsus alternans. The peaks are marked by gray circles, and the lengths between two consecutive peaks are shown. Note that  there are two peaks around the second second, and we {choose} the one with the larger value as the peak. 
(b) The time-frequency (TF) representation of $R(t)$ determined by the synchrosqueezing transform. The dominant curve in the TF representation is associated with the IF induced by the heart rate variability (HRV). 
It is clear that the pulse around {the} $5^{th}$ second takes a longer time to finish, which leads to the slower instantaneous frequency. Indeed, the y-axis of the dominant curve around time $5.3$-th second is $1/1.15=0.87$, which reflects how fast the signal oscillates at that moment. Note that the artifacts around {the} $5.5$ second and {the} $7.5$ second do not play a major role in the analysis {results}.}\label{fig:SSTExample}   \end{figure}

\begin{figure}[h!]
    \begin{center}
    \includegraphics[width=.6\textwidth]{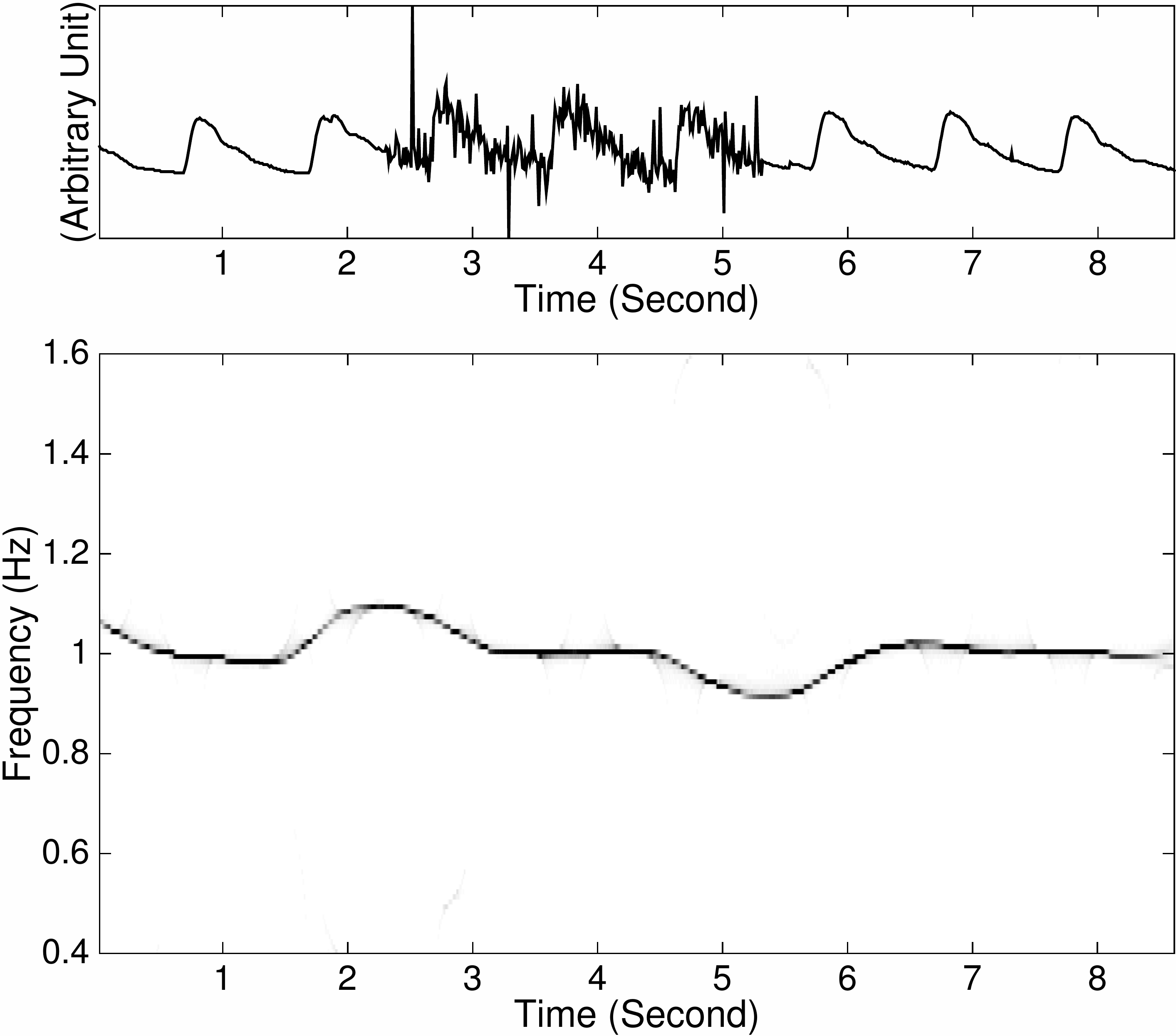}     
    \end{center}
    \caption{(a) The radial pulse signal $R(t)$ recorded from a normal subject {that is} contaminated by the autoregressive and moving averaged noise generated from the student $t_3$ i.i.d. random process from the $2.5$-th second to the $5.5$-th second, which is denoted as $Y(t)$. The radial pulse signal is the same as that shown in Figure \ref{fig:SSTExample}. Note that there are several spikes in $Y(t)$, which are generated {by} the fat tail natural of the student t 3 random variable.
(b) The time-frequency (TF) representation of $Y(t)$ {is} determined by the synchrosqueezing transform (SST). The dominant curve in the TF representation is associated with the IF induced by the heart rate variability (HRV). 
It is clear that the TF representation determined by SST is robust to the noise even if the noise is time-dependent with {a} fat tail distribution, while the dominant curve representing the instantaneous frequency is slightly deformed due to the noise.}\label{fig:SSTExampleNoisy} 
  \end{figure}

Here, we show the results of pulse analysis by applying the SST in Figure \ref{fig:SSTExample}. To demonstrate its benefit on {a} noisy signal, we artificially add noise on the second half of the signal, and show the result in Figure \ref{fig:SSTExampleNoisy}. In this example, the noise is an ARMA(1,1), where ARMA stands for autoregressive and moving averaging, time series determined by the autoregression polynomial $a(z)=0.5z+1$ and the moving averaging polynomial $b(z)=-0.3z+1$, with the innovation process taken as i.i.d. student $t_3$ random variables so that the signal to noise ratio, defined as $20\log\frac{\text{std}(\text{signal})}{\text{std}(\text{noise})}$, is $0$ dB. It is clear that {though} the signal to noise is low, the noise is non-stationary since it exists only over a finite interval, and the noise has the fat tail behavior described by the student $t_3$ random variables. {Thus,} the SST algorithm could reliably extract the instantaneous frequency, so that we could obtain reliable HRV information.

\subsection{Feature extraction by estimation of the wave shape functions}

As discussed in Section \ref{Section:AdaptiveHarmonicModel}, the main feature {of} interest {for} the pulse analysis is the wave-shape function in the adaptive non-harmonic model, in particular in the pulse spectral analysis. In this section, we describe an algorithm to extract the wave shape function.
Note that {since} the wave-shape function can be expanded by its Fourier coefficients, the pulse signal $f(t)=A(t)s(\phi(t))$ can be represented as
\begin{equation}\label{IMTexpansionVersion1}
f(t)=A(t)\sum_{\ell\in\ZZ}\left[\alpha_{\ell}\cos(2\pi \ell\phi(t))+\beta_{\ell}\sin(2\pi\ell\phi(t))\right],
\end{equation}
where $\alpha_{\ell}\in\RR$ and $\beta_{\ell}\in\RR$ are the Fourier coefficients of the shape function $s$. 
In this study, we estimate the wave shape functions based on (\ref{IMTexpansionVersion1}) using the standard functional regression \cite{Ramsay_Silverman:1997,Chui_Lin_Wu:2016}.

Consider the wave shape functions $s$ with parameters $\delta, D,\theta$ in (\ref{decompShape}). To simplify our discussion, we assume that $\theta=0$ and that the noise is stationary, that is $\sigma=1$. We would choose $D=6$ in this study, based on the discussion in Section \ref{Section:methodology}. Thus, the pulse signal (\ref{IMTexpansionVersion1}) becomes
\begin{equation}\label{IMTexpansionVersion2}
f(t)=\, A(t)\alpha_0+A(t)\sum_{\ell=1}^{D} \left[\alpha_{\ell}\cos(2\pi \ell\phi(t))+\beta_{\ell}\sin(2\pi\ell\phi(t))\right]\nonumber.
\end{equation}
After discretization by the sampling period $\Delta t>0$ over time interval $[\Delta t, N\Delta t]$, the recorded pulse signal is saved digitally as a $N$-dim vector, $\vY\in\RR^N$, so that its $l$-th entry is $\vY_l=f(l\Delta t)+\Phi_l$, where $l=1,\ldots, N$ and $\Phi$ is a random vector satisfying $\textup{var}(\Phi_l)=1$ for all $l$, which might not be Gaussian and the covariant matrix might not be the identity.
Denote the discretized estimators of $A(t)$ and $\phi(t)$ by $\widetilde{A}\in\RR^{N}$ and $\widetilde{\phi}\in\RR^{N}$. Note that it has been {well established} the discretized estimation of $A$ and $\phi$ are accurate with error of order $\epsilon$ \cite{Chen_Cheng_Wu:2014}. To {simplify} the discussion, {we} assume that the estimates $\widetilde{A}$ and $\widetilde{\phi}$ are precise without error; that is, $\widetilde{A}(l)=A(l\Delta t)$ and $\widetilde{\phi}(l)=\phi(l\Delta t)$ for all $l=1,\ldots,N$. In the general case, the analysis result will {deviate} by an error of order of $\epsilon$.
We thus construct the following ``functional vectors''  
\begin{equation*}
\boldsymbol{c}=[c_0^T,c_{1}^T,\ldots,c_{D}^T,d_{1}^T,\ldots,d_{D}^T]^T\in \RR^{2D\times N},
\end{equation*}
where $\ell=0,\ldots,D$ and $c_{\ell}, d_{\ell}$ are $N$-dim vector whose $k$-th entries are
\begin{equation}
c_{\ell}(k):=\widetilde{A}(k)\cos(2\pi\ell\widetilde{\phi}(k)),\quad d_{\ell}(k):=\widetilde{A}(k)\sin(2\pi\ell\widetilde{\phi}(k)),\nonumber
\end{equation} 
where $k=1,\ldots, N$. As a result, the recorded pulse signal satisfies
\begin{equation}\label{model:with_shape_and_trend:regression_model}
\vY=\boldsymbol{\gamma}^T\boldsymbol{c}+\Phi
\end{equation}
where $\boldsymbol{\gamma}=[\alpha_0,\alpha_{1},\ldots,\alpha_{D},\beta_{1},\ldots,\beta_{D}]\in\RR^{2D+1}$.

To estimate the parameters $\alpha_{\ell}$ and $\beta_{\ell}$ from the functional vectors $\boldsymbol{c}$, observe that the $(2D+1)\times (2D+1)$ matrix $\boldsymbol{c}\boldsymbol{c}^T$ is diagonal dominant. Since $\mathbb{E}(\Delta t\Phi\boldsymbol{c}^T)=0$ and $\mbox{var}(\Delta t\Phi\boldsymbol{c}^T)=O(L\Delta t)$. 
Thus we can estimate $\boldsymbol{\gamma}$ by
\begin{equation}\label{estimator_shapefunction}
\widetilde{\boldsymbol{\gamma}}:=(\vY\boldsymbol{c}^T)(\boldsymbol{c}\boldsymbol{c}^T)^{-1},
\end{equation}
where $\widetilde{\boldsymbol{\gamma}}=[\widetilde{\alpha}_{0},\widetilde{\alpha}_{1},\ldots,\widetilde{\alpha}_{D},\widetilde{\beta}_{1},\ldots,\widetilde{\beta}_{D}]^T\in\RR^{2D+1}$, 
and hence estimate the oscillatory signal by
\[
\widetilde{f}(t):=\widetilde{\boldsymbol{\gamma}}^T\boldsymbol{c}. 
\]
{When} $\widetilde{f}$ contains an accurate estimation of the wave-shape function $s$ with $L^2$ error of order $\epsilon$, we could take the Fourier coefficients $\widetilde{\boldsymbol{\gamma}}$ into account as the feature of the pulse signal. We call the $2D+1$ dimensional vector $\widetilde{\boldsymbol{\gamma}}$ the {\it spectral pulse signature (SPS)} for the recorded pulse signal.
Note that the $\ell$-th component of the power spectrum of $s$ could be estimated by $(\widetilde{\alpha}^2_{\ell}+\widetilde{\beta}^2_{\ell})/4$. The main benefit of this approach is that the influence of HRV, which is modeled by IF, is eliminated automatically and the wave-shape function is better estimated. 

\subsection{{Comparing the present method to previous ones}}

Here we summarize the difference between our approach and the commonly applied spectral analysis \cite{Taylor:1966,Wei_Chow:1985,LinWang_Jan_Shyu_Chiang_Wang:2004}. Recall that in spectral analysis, the power spectrum of a selected recorded pulse signal is evaluated and the energy of different harmonic modes are considered {as} features of the subject, {indicating aspects of} physiological health. The selection criteria for the pulse signal interval is that the heart rate is almost a constant \cite{LinWang_Jan_Shyu_Chiang_Wang:2004,Hsu_Chao_Hsiu_Wang_Li_Wang:2006}{, since the power spectrum approach is sensitive to the non-constant frequency and non-constant amplitude signal}. The first difference is that based on the adaptive non-harmonic model and SST, the instantaneous frequency and amplitude modulation {can be} accurately {obtained}, so we do not need to select an interval from a recorded pulse signal. {Second, since the HRV and pulsus alternans are physiologically} inevitable, {the proposed} approach is more physiologically feasible. 

Third, note that the information obtained from the spectral analysis is different from the SPS. Indeed, under the assumption for (\ref{IMTexpansionVersion2}), we have the following direct expansion:
\[
s(t)=\sum_{\ell=-D}^D\hat{s}(\ell)e^{i2\pi \ell t}=\alpha_0+\sum_{\ell=1}^D\left[\alpha_\ell\cos(2\pi \ell t)+\beta_\ell\sin(2\pi \ell t)\right],
\]
where $\alpha_0=\hat{s}(0)$ and for $\ell=1,\ldots,D$, $\hat{s}(\ell)=a_\ell e^{i\theta_\ell}$, $a_\ell\geq 0$, $\theta_\ell\in[0, 2\pi)$, $\alpha_\ell=2a_\ell \cos(\theta_\ell)$ and $\beta_\ell=-2a_\ell\sin\theta_\ell$. Here, $\theta_\ell$ reflects the phase of the $\ell$-th harmonic component hidden inside the wave shape signal. Clearly, since power {spectrum} analysis {commonly takes} $|\hat{s}(\ell)|^2=a_\ell^2$, the phase information of the $\ell$-th harmonic component is missed. {However, in SPS the phase information is preserved and the hemodynamics could be more faithfully captured.}

\section{Testbed: the radial pulse diagnosis}\label{Section:materialresult}

To test how well the proposed adaptive non-harmonic model and the algorithm work in practice, in this section we study the radial pulse wave signal on congestive heart failure subjects. 
The signals are recorded from three landmarks. First, the radial styloid process; second the middle position between the radial styloid process and the palm and the proximal point with the same distance from the first to the second location.
We mention that these three locations are commonly recognized in the pulse diagnosis; the first one is called {\em guan}, the second one is called {\em chun} and the third one is called {\em chi}. {We} thus use guan, chun and chi to refer to these three locations in this study. 
\footnote{In the literature sometimes chun is called {\em inch} or {\em chon}, guan is called {\em bar} or {\em gwan} and chi is called {\em cubit} or {\em cheok}.} 
  
\subsection{Material}

All protocols in this study {were} approved by the Institutional Review Board of Chang Gung Memorial Hospital, Linkou, Taoyuan, Taiwan (93-6288), Taiwan and written informed consent was obtained from {all} patients. 
Nineteen normal subjects {without history of} heart disease are included in the control group, and $17$ subjects with congestive heart failure (CHF) are included in the study group. The diagnosis of CHF subjects is based on the criteria indicated by the Framingham heart study. The participants were invited for pulse examination in a quiet and temperature-controlled room in Chang Gung Memorial Hospital, Linkou branch, Taiwan. Pulse wave signals were recorded from chun, guan and chi positions of both hands by a tonometer (Wang's sphygmometer, PDS-2000). The sampling rate of the signal is at $100$Hz. For each subject, we collect $10$ seconds signal for each position on both {hands}, and repeat $2$ or $3$ times. The pulse wave was recorded in sitting position with the wrist comfortably {resting} on a small pillow at the level of {the} heart

We recruited $17$ patients with CHF for the study group and $19$ normal individuals for the controls. 
The age of the study and control group are $64.3\pm 23.7$ and $63.2\pm 15.8$ respectively. The male/female ratio were $15/7$ in the study group and $10/10$ in the control group. There were no significant {differences} in age {or} sex. As a result, we obtain $53$ (respectively $39$) pulse wave signals recorded from chun from the normal (respectively CHF) group, $55$ (respectively $34$) pulse wave signals recorded from guan from the normal (respectively CHF) group and $50$ (respectively $41$) pulse wave signals recorded from chi from the normal (respectively CHF) group.

\subsection{Statistical Analysis and Global pulse signature (GPS)}

To test if the SPS indices of the normal group {differ from} those of the CHF group, we apply the currently developed one-way ANOVA for functional data, called the globalized pointwise F (GPF) test \cite{Zhang_Liang:2013}. For readers having interest in this technique, we refer them to \cite{Zhang_Liang:2013}. In this study, we {consider} p values less than $0.01$ as significant.

As the SPS index is a high dimensional vector, it is not easy to visualize. To provide an easy-to-use index for the pulse diagnosis, we consider the following approach to integrate the information in the SPS index. Suppose from a fixed position of the $i$-th subject we obtain a SPS index $\widehat{\boldsymbol{\gamma}}_i$. The associated outcome of this SPS index is denoted as $y_i$, and $y_i=1$ (respectively $y_i=0$) means the subject is in the CHF group (respectively control group). Thus we have the dataset $\{\widehat{\boldsymbol{\gamma}}_i,y_i\}_{i=1}^N$. {When} the SPS index is located in the $2D+1$ Euclidean space and the sampling size is limited, we apply the partial least squares (PLS) regression to find a linear regression model by projecting SPS and the response variables to a new space. Here we briefly recall the PLS regression. PLS regression finds components from $\mathcal{X}=\{\widehat{\boldsymbol{\gamma}}_i\}_{i=1}^N$ that are relevant for the outcome $\mathcal{Y}=\{y_i\}_{i=1}^N$ by seeking a set of components that performs a simultaneous decomposition $\mathcal{X}$ and $\mathcal{Y}$ with the constraint that these components explain as much as possible of the covariance between $\mathcal{X}$ and $\mathcal{Y}$. Then the decomposition of $\mathcal{X}$ is applied to predict the group.
For details about PLS regression, see \cite{Rosipal_Kr:2006}. 
Suppose the PLS regression coefficient is $\beta\in \RR^{(2D+1)\times 1}$. Then the prediction result under the linear regression model for $\widehat{\boldsymbol{\gamma}}_i$, denoted as $\hat{y}_i:=[1 \,\,\,\widehat{\boldsymbol{\gamma}}_i]\beta\in\RR$, is referred to {as} the {\it global pulse signature (GPS)} index. The GPS index is integrated information derived from the SPS index, which reflects the subject's condition {of} interest. 

For the {purpose of prediction} based on GPS, we {can} further apply the receiver operating characteristic (ROC) to determine the threshold to classify the subjects into two groups. We report the sensitivity, specificity, accuracy and the area under curve (AUC) to evaluate the classification result. The confidence interval (CI) of AUC is evaluated by $1000$ bootstrap replicas. To assess how the results based on PLS and ROC will generalize to an independent data set, we run leave-one-out cross validation (LOOCV) $200$ times for each position of both hands, and report the accuracy.

\subsection{Results}

\begin{figure}[h!]
    \begin{center}
    \includegraphics[width=.78\textwidth]{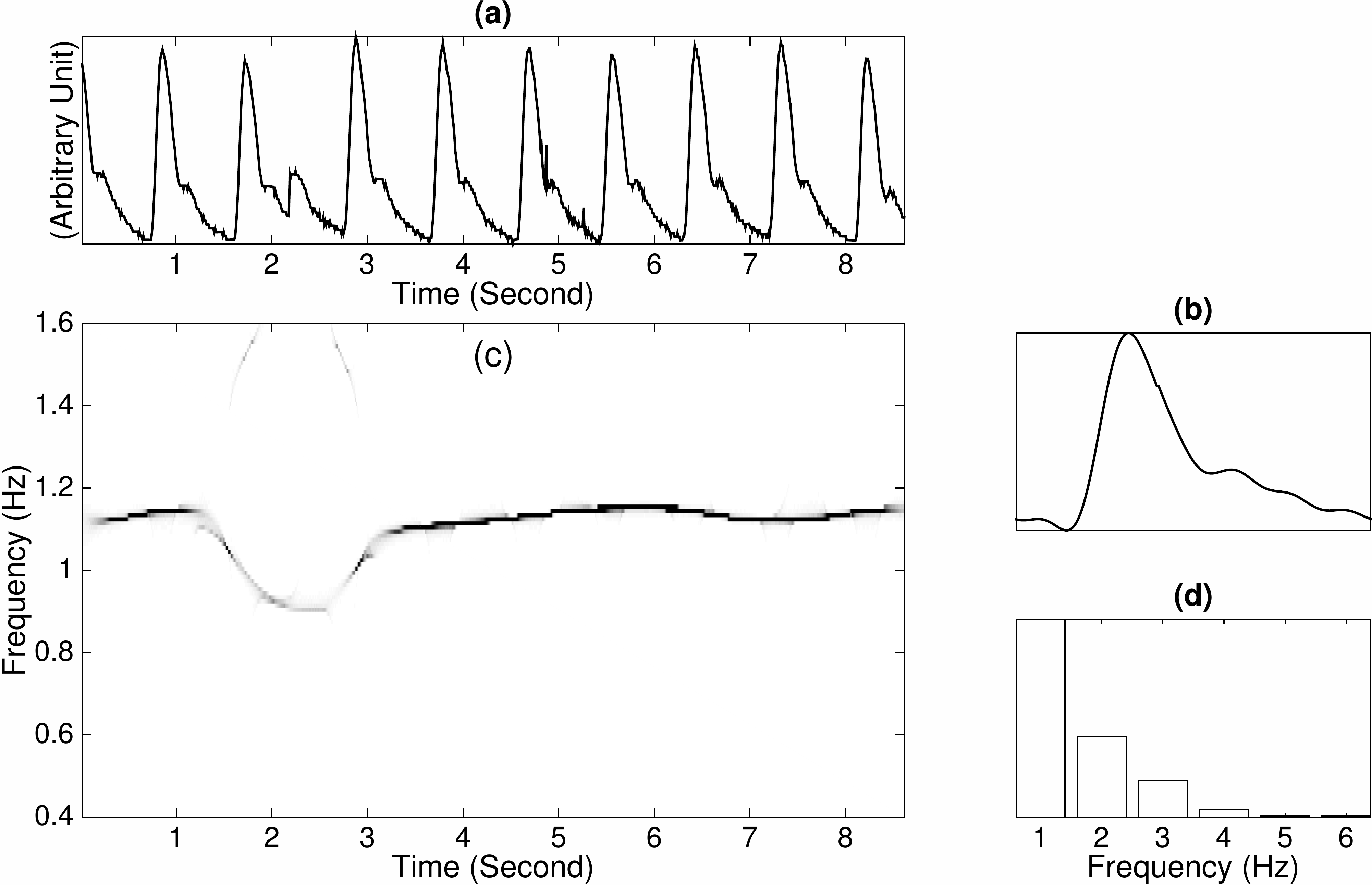}     
    \end{center}
    \caption{(a) The pulse signal recorded from a subject with congestive heart failure (CHF), {with the} y-axis {as} the arbitrary unit. (b) {The} estimated wave shape function for the subject with CHF. 
(c) {The} time-frequency (TF) representation determined by the synchrosqueezing transform. It is clear that the pulse around {the} $2.5^{th}$ second takes a longer time to finish, which leads to the slower instantaneous frequency (the dominant curve in the TF representation is around 0.9 Hz at time$ 2.5^{th}$ second). It is also clear that the instantaneous frequency is not constant due to the inevitable heart rate variability (HRV).
Note that while there is a significant deviation at the $2.5$-th second, the estimation result catches most of the shape information. (d) {The} power spectrum of the estimated wave shape function shown in (b). 
}\label{fig:ShapeExtractionExample} 
  \end{figure}

First, we show the synchrosqueezing transform of the pulse wave signal from a subject with CHF and the associated estimated wave shape function in Figure \ref{fig:ShapeExtractionExample}. Note that due to the inevitable deviation, for example the one at the $4$-th second, and the HRV, the power spectrum estimated from the pulse wave signal is spreading. Note that estimating the wave shape function could be viewed as the power spectrum analysis of the pulse wave signal after correcting the IF and AM. In other words, as we could estimate IF and AM accurately from the pulse wave signal, we could resample the signal according to the estimated IF and then normalize the signal by the estimated AM.

\begin{figure}[h!]
    \begin{center}
    \includegraphics[width=.95\textwidth]{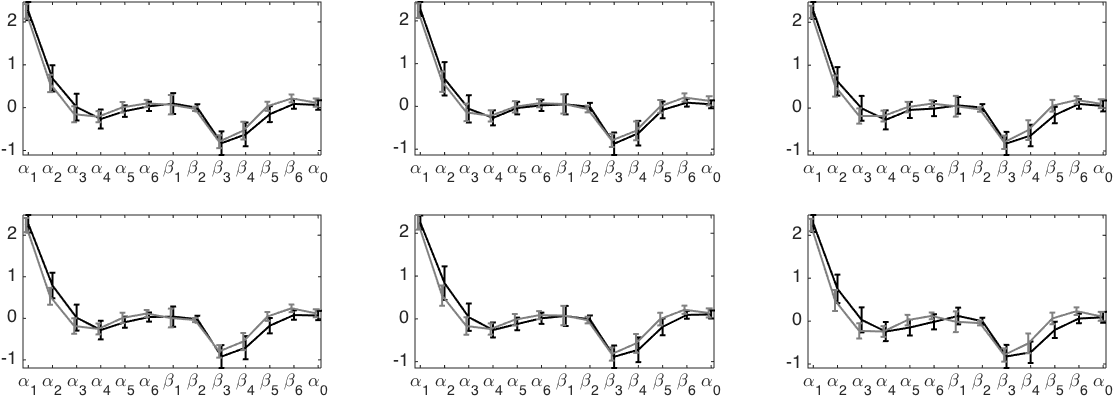}     
    \end{center}
    \caption{The deviation from the mean of each entry of the spectral pulse signature (SPS) is shown in the error bar plot. The error bar is determined by the standard deviation. In the top row, the distribution of SPS determined from guan, chun and chi {positions} of the left hand are shown from left to right. In the bottom row, the distribution of SPS determined from guan, chun and chi of the right hand are shown from left to right. The results of the normal subjects are shown in the gray error bar, and those of the subjects with congestive heart failure are shown in the black error bar. }\label{fig:SPSErrorBar}
  \end{figure}

The set of SPS indices of different groups from different positions {is} shown in Figure \ref{fig:SPSErrorBar}. We could see that the means of the normal group and the CHF group are different. The GPF functional ANOVA test shows that the SPS indices evaluated from all positions on both hands are significantly different. Except {for} the p value of {the} chun position on the right hand, which is $4.4\times 10^{-4}$, the p values of other positions are $<10^{-4}$.

\begin{figure}[h!]
    \begin{center}
    \includegraphics[width=.9\textwidth]{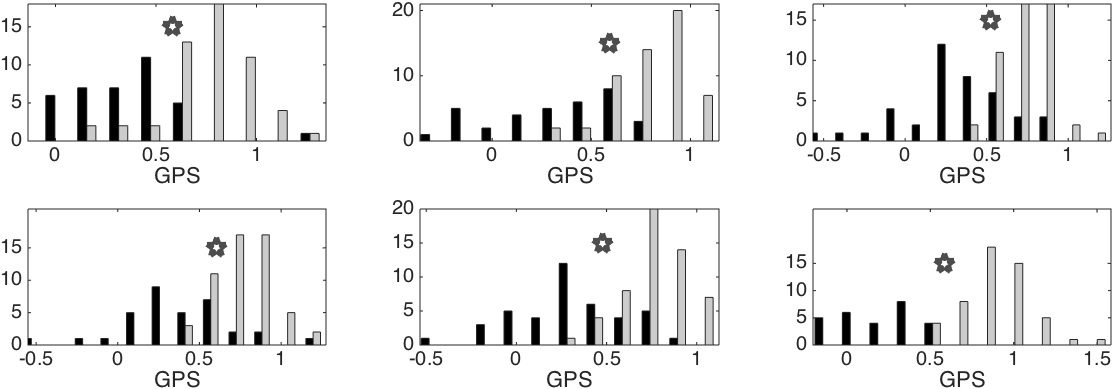}     
    \end{center}
    \caption{The histograms of GPSs determined from chun, guan and chi positions on the left (respectively right) hand are shown on the left, middle and right subfigures on the top (respectively bottom) row. The normal group is shown in the gray color and the congestive heart failure group is shown in black. {Gray} stars represent the determined thresholds by the ROC binary classification.}\label{fig:GPS}
  \end{figure}

\begin{figure}[h!]
    \begin{center}
    \includegraphics[width=.3\textwidth]{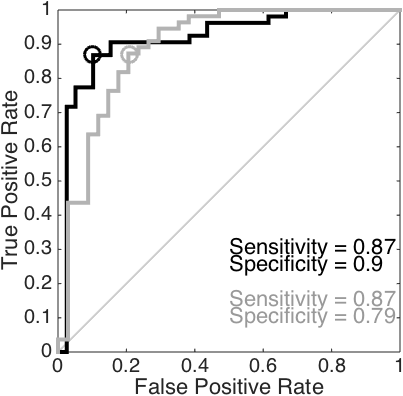}     
    \includegraphics[width=.3\textwidth]{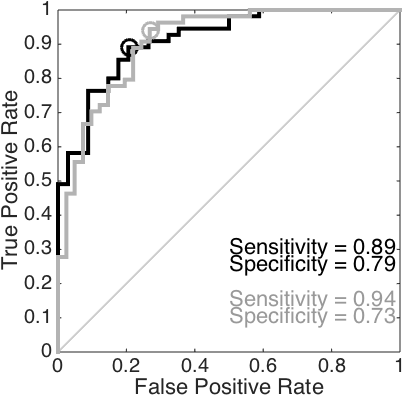}      
    \includegraphics[width=.3\textwidth]{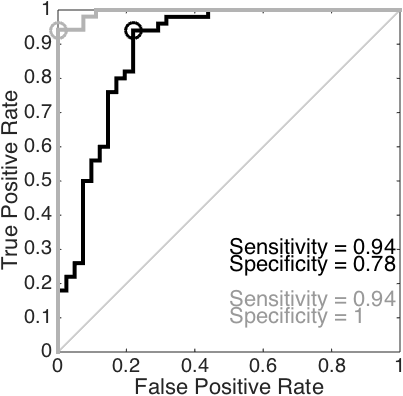}      
    \end{center}
    \caption{Left (respectively middle and right): the receiver operating curve (ROC) of the global pulse signature (GPS) of the chun (respectively guan and chi) position on the left hand is shown in the black curve and that of the right hand is shown in the gray hand. {Black} and gray circles are the optimal operating point of the ROC.}\label{fig:ROC}
  \end{figure}

Then we apply PLS to obtain the GPS index to distinguish the two groups. The histogram of GPS indices determined from different positions on both hands are shown in Figure \ref{fig:GPS}. It is clear that the GPS of subjects in the CHF group is smaller than that of the control group. The ROC analysis results, including the sensitivity and specificity and AUC, from different positions are shown in Figure \ref{fig:ROC}. 
The sensitivity, specificity, accuracy, AUC and the accuracy of LOOCV of the GPS determined from different positions on both hands are summarized in Table \ref{tab:allROC}. 

\begin{table}[ht!]
\centering
\begin{tabular}{|c|c|c|c|c|c|}\hline
\multirow{2}{*}{} &\multicolumn{5}{c|}{Left hand}  \\\cline{2-6}
       &SEN      &SPE  & AC  &AUC  &LOOCV  \\\hline
chun & 0.87    & 0.9  & 0.88 & 0.92 (0.81-0.96) & 0.8   \\\hline
guan  & 0.89    & 0.79  & 0.85 & 0.91 (0.85-0.96)& 0.74 \\\hline
chi & 0.94    & 0.78 & 0.87 & 0.89 (0.8-0.95)  & 0.74 \\\hline\hline
\multirow{2}{*}{} &\multicolumn{5}{c|}{Right hand}  \\\cline{2-6}
       &SEN      &SPE  & AC  &AUC  &LOOCV  \\\hline
chun  & 0.87      & 0.79 & 0.84 & 0.89 (0.79-0.95) & 0.74   \\\hline
guan  & 0.94     & 0.73 & 0.85 & 0.91 (0.83-0.95)  &  0.79   \\\hline
chi  & 0.94   & 1   & 0.96 & 1 (0.97-1)  & 0.94 \\\hline
\end{tabular}
\caption{{Sensitivity} (SEN), specificity (SPE), accuracy (AC), the area under curve (AUC) and the accuracy of the leave-one-out cross validation (LOOCV) of the global pulse signature evaluated from different positions on both hands. The confidence {intervals} of the AUC {are} reported in {parentheses}.}
\label{tab:allROC}
\end{table}

\section{Discussion}\label{Section:discussion}

In summary, in this paper we study the physiological signal by the adaptive non-harmonic model, the SST and the functional regression technique. The usefulness of the proposed scheme is supported by an encouraging analysis result of the radial pulse signal. 
Although we analyze the radial pulse signal as the test {case for} this study, {it should be noted} that the proposed model and analysis technique could be applied to other pulse signals obtained by different instruments. For example, contact photoplethysmogram (PPG) measurement \cite{Allen:2007} or PhysioCam non-contact PPG measurement \cite{Davila:2012Thesis}, which represent the changes of blood volume in the vessel obtained through an optical transmission measurement or real-time camera images, and hence reflect the change in vascular blood volume associated with the cardiac beat.
While these signals represent different {aspects} of the hemodynamics {than} the radial pulse signal we study in this paper, we could expect to obtain a broader angle of view about human health if information obtained from these signals {could be combined}. We {will} report the {research} progress in the future work.
Since the potential of the SST and other nonlinear TF analysis techniques have been shown in this study and other clinical problems, for example, \cite{Lin_Wu_Tsao_Yien_Hseu:2014,Baudin_Wu_Bordessoule_Beck_Jouvet_Frasch_Emeriaud:2014,Wu_Talmon_Lo:2015}, by taking the wave-shape model into account, we could expect a broader application and better analysis {results}.

\begin{figure}[h!]
    \begin{center}
    \includegraphics[width=.6\textwidth]{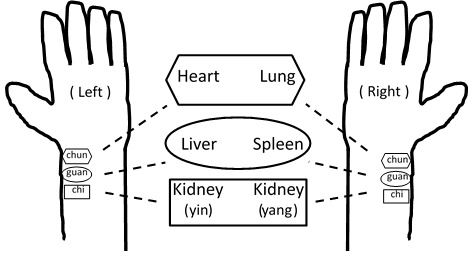}     
    \end{center}
    \caption{An illustration of anatomical locations associated with the terminologies used in pulse diagnosis. In the TCM pulse diagnosis, it is stated that the pulse on the right hand manifests the condition of Qi \cite{Leung:2011,Wiseman_Ye:2014}, while the pulse on the left hand reflects the blood. For different pulse positions, left chun, guan and chi reflect the heart, the liver, and the kidney respectively; the right chun, guan and chi reflect the lung, the spleen, the kidney respectively. The left chi manifests mainly the kidney yin, and the right chi manifests mainly the kidney yang (the Life Gate in TCM). 
}\label{fig:ChunGuanChi} 
  \end{figure}

We {also point out} the relationship of the pulse signal analysis results with the {traditional Chinese medicine} (TCM) theory. In short, the results we obtained in this study {could} lead to a potential {means} to help establish the foundation of TCM theory. See Figure \ref{fig:ChunGuanChi} for an illustration of the summary of the pulse diagnosis theory in the TCM theory. 
Note that while TCM has been widely applied in the eastern culture, up to now, a systematical/scientific theory understanding the mechanism beyond pulse diagnosis is not yet well established \cite{Leung:2011}, but its usefulness {has been demonstrated} \cite{Oparil_Joseph:2006}. There are several {studies} based on hemodynamics aiming to understand the mechanism; for example, Wang et. al. \cite{Wang_Hsu_Jan_Wang:2010,Wang_Wang_Jan_Wang:2012} proposed that the pulse wave consists of numerous harmonic waves, and each harmonic wave is associated with an internal organ and carries information of different meridians over the body. {That study also} noted that the harmonic waves of the upper, middle and lower section of the body may correspond to the three sections of pulse in timeline and that the waveform on both hands are not totally the same. On the other hand, our approach is purely from the phenomenological viewpoint and adaptive harmonic analysis.
In our results, the classification result of the CHF group and the normal group is better on the right hand side, especially the right chi. Since the right chi position {demonstrates} the most significant difference between groups, it may reveal the importance of {the} right chi (kidney yang) signal in evaluating the clinical condition of CHF. According to TCM theory, kidney yang is the fundamental support to maintain the human life. In {Western} medicine, the hemodynamics of renal artery, such as the blood pressure, plays an important role {in regulating the overall} whole cardiovascular function. On the other hand, a possible cause for this observation is that the pulse pressure of the right arm is usually higher than the pulse pressure of the left arm. It is presumed that the pulse wave signal of the right arm may have a higher signal-to-noise ratio than the pulse wave signal of the left arm.
The left chun and right guan position are the two second important roles in our analysis. According to TCM theory, {the} left chun position (heart) manifests the general condition of the cardiovascular system, hence our results {are consistent with} clinical experience. 
Thus, our {findings} partially support the pulse diagnosis theory in TCM that the waveforms on different positions of radial artery contain different information. {A} similar finding {was} reported {by} Young et. al., {who} found that the although the augmentation indices determined from the radial pulse waves recorded from chun, guan and chi are not significantly different, the estimated aortic augmentation indices determined from the radial pulse waves recorded from chun, guan and chi are not identical \cite{Jeon_Kim_Lee_Lee_Ryu_Lee_Kim:2011}. 
In conclusion, since the study of pulse diagnosis from different aspects is an active field, we would expect our proposed model and method could help to further study the experiences and working practice {of} pulse diagnosis, e.g. the nature and dynamic of disease, and its relationship to modern hemodynamics.

Limitations of this study should also be mentioned.
First, the tonometer (PDS-2000) we applied in this study records only the two-dimensional data (pressure-time) of the pulse. Since {more advanced} instruments {now available can} obtain three dimensional data \cite{Luo_Chung_Yeh_Si_Chang_Hu_Chu:2012}, further study should be {undertaken}. 
Second, although the phenomenological model we propose is capable of capturing IF and AM as well as the wave shape function, it is clearly not the optimal solution. It is clear that there is a more {complex} interaction between IF, AM and the wave shape function than what we consider in the adaptive non-harmonic model. On the one hand, a more general model, like the time-varying wave shape function, could be considered based on the physiological background. On the other hand, we conjecture that this relationship might be better captured by combining the existing hemodynamic models with the proposed phenomenological model. This finer model might {better} capture the physiological information hidden inside the pulse wave signal and lead to a better algorithm to better study the recorded pulse wave signal. A systematic study and its application of this issue will be reported in future work. 
Moreover, from the clinical viewpoint, the sample size in this study is limited, and the interesting clinical problems, like outcome prediction or early CHF diagnosis are not discussed. Also, to simplify the discussion and avoid possible confounders, in this study we limit our analysis to subjects with CHF. Thus, we could not make the final conclusion about the pulse diagnosis. To use the research results in clinics, a larger scale clinical study with CHF and other diseases is needed to conclude the {current findings}, and we will report our continuing {research} in the future.

\section{Acknowledgement}
%
H.-T. Wu also thanks Professor Jin-Ting Zhang for sharing the GPF code. Part of this work was done during H.-T. Wu's visit to National Center for Theoretical Sciences, Taiwan, and he would like to thank NCTS for its hospitality.

\bibliographystyle{vancouver}
\bibliography{../bib/Optimization,../../tex20130723/bib/MathBooks,../bib/MedicalBooks,../bib/PulseDiagnosis,../bib/PhysiologicalVariability,../bib/noisyManifold,../bib/TFanalysis,../bib/Thesis,MultivariateTimeSeries}

\end{document}